%% file: main.tex
\documentclass[sigconf]{acmart}

\usepackage{xspace}
\usepackage{graphicx} 
\usepackage{subcaption}
\usepackage{hyperref}
\usepackage{url}
\usepackage{algorithm}
\usepackage{algorithmic}
\usepackage{multirow}

\newif\ifcomments
\commentstrue
\usepackage{enumitem}
\setlist{topsep=1pt, leftmargin=*}

\newcommand{\ourframework}{TBNet\xspace}
\newcommand{\secure}{$M_{T}$\xspace}
\newcommand{\normal}{$M_{R}$\xspace}

\AtBeginDocument{%
  \providecommand\BibTeX{{%
    \normalfont B\kern-0.5em{\scshape i\kern-0.25em b}\kern-0.8em\TeX}}}

\copyrightyear{2024}
\acmYear{2024}
\setcopyright{rightsretained}
\acmConference[DAC '24]{61st ACM/IEEE Design Automation Conference}{June 23--27, 2024}{San Francisco, CA, USA}
\acmBooktitle{61st ACM/IEEE Design Automation Conference (DAC '24), June 23--27, 2024, San Francisco, CA, USA}
\acmDOI{10.1145/3649329.3658251}
\acmISBN{979-8-4007-0601-1/24/06}

\begin{document}

\title{\ourframework: A Neural Architectural Defense Framework Facilitating DNN Model Protection in Trusted Execution Environments}

\author{Ziyu Liu$^1$, Tong Zhou$^1$, Yukui Luo$^2$, Xiaolin Xu$^1$}
\email{
{liu.ziyu4, zhou.tong1, x.xu}@northeastern.edu, yluo2@umassd.edu}
\affiliation{
\institution{{$^1$Northeastern University, $^2$University of Massachusetts Dartmouth}}
\country{}
}

\renewcommand{\shortauthors}{Z.Liu et al.}

\begin{abstract}
Trusted Execution Environments (TEEs) have become a promising solution to secure DNN models on edge devices. 
However, the existing solutions either provide inadequate protection or introduce large performance overhead. Taking both security and performance into consideration, this paper presents \ourframework, a TEE-based defense framework that protects DNN model from a neural architectural perspective. Specifically, \ourframework generates a novel \underline{T}wo-\underline{B}ranch substitution model, to respectively exploit (1) the computational resources in the untrusted Rich Execution Environment (REE) for latency reduction and (2) the physically-isolated TEE for model protection. 
Experimental results on a Raspberry Pi across diverse DNN model architectures and datasets demonstrate that \ourframework achieves efficient model protection at a low cost.

\end{abstract}
\maketitle

\section{Introduction}

Deep neural network (DNN) models are widely deployed on edge devices for diverse applications \cite{murshed2021machine}. These on-device DNNs benefit from local model execution to reduce inference time. Moreover, on-device DNN mitigates the risk of data leakage, as sensitive data predominantly resides on the edge \cite{isakov2019survey}. Despite these advantageous aspects of deploying DNNs on edge, it introduces security concerns. These well-trained models represent valuable intellectual property (IP), and the edge deployments expose them to model stealing attacks, in which adversaries can extract the model architecture and weights, using illegal memory access \cite{sun2021mind}, side-channel attacks \cite{batina2019csi}, or direct interactions with the model \cite{jagielski2020high}. These attacks pose an emerging threat and may result in potential financial losses for model vendors.

Recently, the Trusted Execution Environments (TEEs) (e.g., ARM TrustZone \cite{armtrustzone}) are employed to secure edge DNN against model stealing attacks \cite{vannostrand2019confidential}. Although TEE ensures the confidentiality of DNN model and computation, its limited memory and computational resources, compared to the untrusted Rich Execution Environment (REE), make it inefficient to deploy the entire model within TEE \cite{chen2019deepattest}. There have been attempts to alleviate the storage and computation burden inside TEE while ensuring DNN model protection \cite{mo2020darknetz,liu2023mirrornet,IP_Protection_TinyML}. These approaches, however, expose certain either well-trained layers or valuable DNN architectures to insecure memory, which can be possibly used by attackers to derive high-performance models, e.g., by fine-tuning valuable DNN architecture with enough data.

Generally, a TEE-base defense should achieve both secure and efficient deployment. Specifically, such a defense method should satisfy four critical requirements: \textbf{(1) preventing attackers from training/ fine-tuning the exposed partial models in REE to achieve a model with comparable accuracy to the victim model} and \textbf{(2) preventing attackers from reverse-engineering deployed layers in TEE} (e.g., infer architecture/computation inside TEE based on exposed information in REE). Simultaneously, the deployed model should \textbf{(3)  preserve the accuracy of the victim model} and \textbf{(4) possess optimized computation and storage in TEE for efficient deployment.}

To meet these requirements, we propose \ourframework, a novel architectural defense framework that generates a \underline{T}wo-\underline{B}ranch substitution model for secure deployment. In particular, \ourframework only includes layer-wise connections from REE to TEE, but not vice versa. In this way, \ourframework not only secures the critical architectural model confidentiality in TEE, but also enjoys the rich computation resources in REE for preserving performance. To achieve these goals, \ourframework transfers partial knowledge from a well-trained victim model into the branch deployed in TEE (i.e., the secure branch). Meanwhile, the branch with partial knowledge remains in REE (i.e., the unsecured branch) to compensate for knowledge lost in the secure branch, preserving the accuracy of victim models. \ourframework employs iterative two-branch pruning to protect the architecture of the victim model and further ease the storage and computation burden of TEE. To prevent the leakage of architectural model confidentiality in TEE through the unsecured branch, \ourframework generates architectural difference between the secure and unsecured branches through rollback finalization. We make the following contributions in this paper: 
\begin{itemize}
    \item We propose \textit{\ourframework}, a TEE-based defense framework for DNN model. \ourframework generates a novel two-branch substitution for secure and efficient deployment of victim models. This architecture defends against model stealing by preventing reverse-engineering of TEE layers and blocking attempts to train exposed partial models in REE to match the accuracy of the victim model.
    
    \item We employ knowledge transfer while optimizing storage and computation overhead in TEE, to preserve the performance of the victim model. By strategically distributing knowledge between two branches, our method achieves a balance between maintaining model accuracy and efficiently utilizing TEE resources.
    
    \item We evaluate \ourframework on a Raspberry Pi across different model architectures and datasets. The experimental results demonstrate that \ourframework significantly reduces TEE memory usage by up to 2.45$\times$ and lowers inference latency by up to 1.22$\times$ compared to fully executing victim models in TEE. 
    
\end{itemize}

\section{Background and Related Works}

\subsection{Trusted Execution Environment}

A Trusted Execution Environment (TEE) is a hardware enclave that guarantees the confidentiality and integrity of data and computations inside \cite{mo2022sok}. It is accomplished through the isolation of this secure area on the main processor, employing a combination of hardware and software methods to ensure its protection. TEE possesses its own dedicated memory area and resources, functioning independently from the standard operating system. 
This design helps guarantee that untrusted users or applications are unable to access the content of the TEE. Standard operations occur in REE, while security-sensitive operations are reserved for TEE. Although TEE is promising in DNN model protection, it is inherently constrained by the memory and computational resources of edge devices that further impact the efficiency.

\subsection{Threat Model}\label{sec:threat_model}
Without loss of generality, we adopt the threat model in related works \cite{batina2019csi,jagielski2020high,liu2023mirrornet}, and elaborate its definition from the attacker and defender perspectives. 
\textbf{Attackers.} Their goal is to obtain a model with comparable performance to the victim model. An attacker can use different attacking methods \cite{batina2019csi,jagielski2020high}, e.g., illegal memory access \cite{sun2021mind}, to extract the model architecture and weights in REE. Following the previous work \cite{mo2020darknetz}, we consider TEE as a secure area and the data, code, and computation within it remain as a black box for attackers. We assume a strong attacker who can extract everything from REE, such as model parameters, computation, and even data transfer activities. Practically, we assume the original victim model from the model vendor is highly optimized. 
\textbf{Defenders.} They aim to ensure the confidentiality of DNN models while maintaining their performance. Specifically, the deployed model should perform well for users but prevent attackers from stealing. To achieve the performance-security balance, defenders can leverage TEE that is widely available on edge devices, such as ARM TrustZone~\cite{armtrustzone}. 

\subsection{Limitations of Prior Arts}\label{sec:limits_of_prior_works}

TEE has been envisioned as a promising solution for DNN model protection, attributed to its secure memory and execution environment \cite{vannostrand2019confidential,mo2020darknetz,sun2023shadownet}. 
Prior works (e.g., \cite{vannostrand2019confidential}) proposed to partition the model and execute the inference sequentially to allow the model to be fully executed inside TEE, thus achieving full protection against model stealing. 
The drawback of this approach is that the overall inference latency is increased, because of the limited computation resources inside TEE. 
As an improvement, DarkneTZ \cite{mo2020darknetz} employs a strategy of partitioning model layers, only executing some sensitive layers within TEE and the remaining layers in the REE. However, the protection of model confidentiality is compromised by the exposure of certain convolution layers to attackers in plaintext format, as it reduces the effort and workload for the retraining process \cite{xiao2019fast}.
At the same time, the intermediate results transmitted between REE and TEE raise concerns.
Specifically, the exposed well-trained layers produce detailed feature maps, serving as the input for the sensitive layers within the TEE. The feature maps generated by these TEE layers are then transmitted back to REE and decrypted in plaintext to execute. By closely monitoring both the input and output of these sensitive layers within the REE, attackers can systematically build and train substitute layers to replace the corresponding layers in the TEE with similar functionality, finally achieving model stealing.
ShadowNet \cite{sun2023shadownet} transforms the weights of linear layers and outsources the intensive computation to untrusted hardware accelerators, subsequently restoring the result inside TEE. However, such a linear transformation approach is vulnerable to strong attackers, who can continuously monitor memory access patterns, match pairs of linear transformations and thereby recover the weights \cite{zhang2022teeslice}. A recent work MirrorNet \cite{liu2023mirrornet} proposes to execute a copy of the victim model in TEE to protect its confidentiality, which employs one-way communication to prevent the leakage of the computation results inside TEE. In this way, the data transmission turns out to be more secure, i.e., only being sent from REE to TEE. However, this solution fails to protect the original DNN architecture, which also represents a valuable model IP \cite{lou2022ownership}. 

\section{Proposed Framework: \ourframework} \label{flow}

\begin{figure*}[t]
  \centering
  \includegraphics[width=0.89\textwidth]{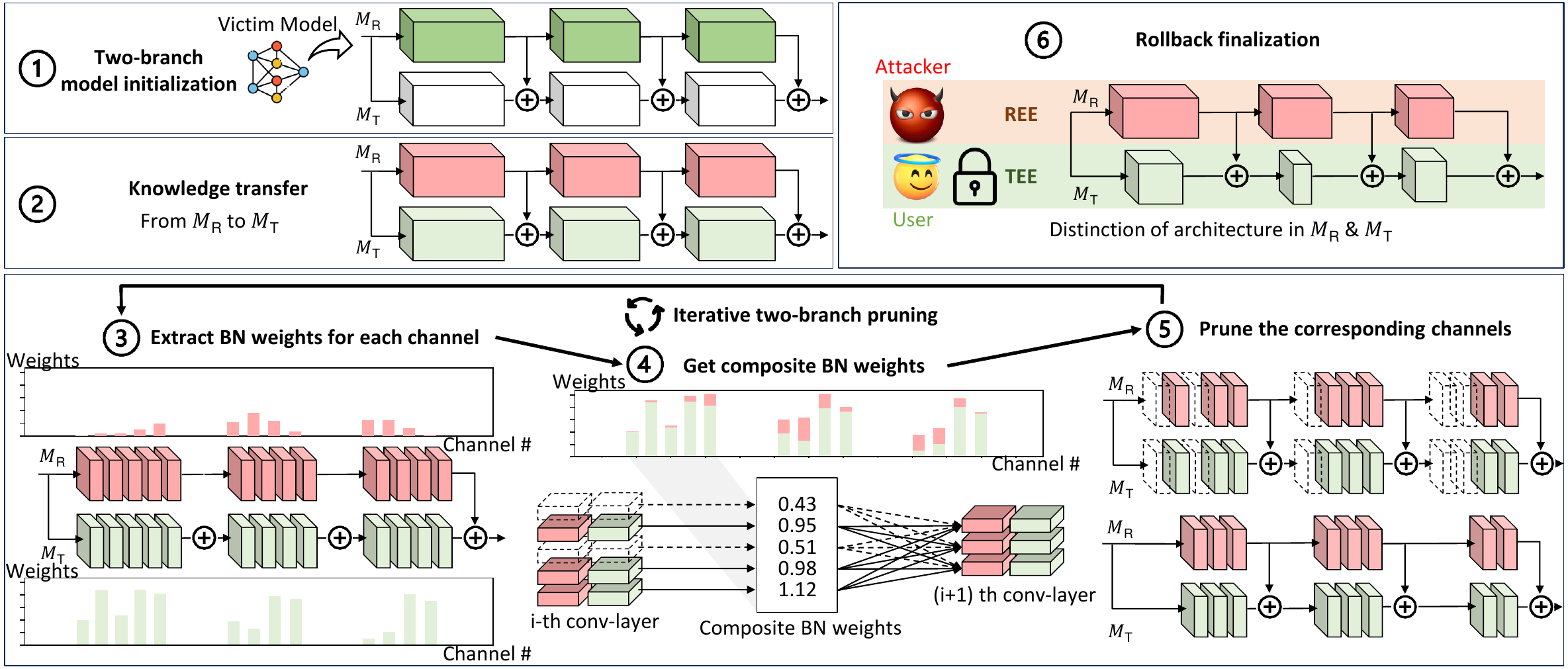}
  \caption{Workflow of the framework. Step \textcircled{1}: \ourframework takes the victim model as the unsecured branch (\normal), then initializes a secure branch (\secure) that has the same model architecture as \normal. Step \textcircled{2} transfers the knowledge of the victim model (i,e., now \normal) to \secure. Step \textcircled{3}-\textcircled{5} apply iterative two-branch pruning to hide the architecture of the victim model and to obtain a lightweight \secure to ease storage and computational burden in TEE. Step \textcircled{6} employs rollback to introduce an architectural distinction between \normal and \secure.}
  \label{fig:overview}
\end{figure*}

\subsection{Design Requirements}

From analyzing the limits of existing works, as well as the generic challenges of using TEE for DNN model protection, we summarize the following design requirements for a TEE-based defense solution:
\begin{itemize}
    \item \textbf{Performance-preserving and optimization in TEE}.
    As a generic requirement for any security solutions, a defense framework should preserve the performance (e.g., inference accuracy) of a victim DNN model. Also, a defense framework can employ both the rich computational resources in REE and the physically secured TEE, to jointly conduct the DNN model inference. Considering the stringent resources in TEE, an efficient defense solution should optimize the model in TEE by reducing its memory usage and computation workload for better hardware efficiency. 

    \item \textbf{One-way context switch between REE and TEE}. 
    While utilizing TEE and REE together to execute a DNN model, a defense solution should address bidirectional intermediate data (e.g., feature map) transmission. Specifically, a defense solution should enforce the one-way context switch from REE to TEE, to prevent the possible model reconstruction using such data leakage.

    \item \textbf{Reduced model confidentiality exposure}. 
    The sub-model executed in REE can leverage rich storage and computational resources, but also leak confidentiality (e.g., its architecture and weights). Therefore, a robust defense strategy must aim to thwart attackers from directly utilizing or fine-tuning the exposed sub-model to attain comparable performance to the victim model. Additionally, it should prevent attackers from deducing the architecture within TEE based on the sub-model in the REE, safeguarding the integrity and confidentiality of the deployed model.
\end{itemize}

To satisfy these design requirements, we propose \ourframework, an architectural defense framework facilitating DNN model protection in TEE, by generating a novel \underline{T}wo-\underline{B}ranch model substitution. In particular, the newly generated model has a lightweight portion within TEE and can achieve similar performance to the victim model. 
Fig. \ref{fig:overview} illustrates the workflow of \ourframework, which includes 6 steps, detailed as follows.

\subsection{Step \textcircled{1}: Two-branch Model Initialization}\label{initial}

We enforce one-way data transmission to avoid leaking the intermediate feature maps generated in TEE. 
In order to achieve this, we initialize a two-branch model by taking the victim model as 
the unsecured branch executed in REE (denoted by \normal), then constructs the secure branch executed in TEE (denoted by \secure) with the exact same architecture as the \normal. 
The reason we employ the same architecture in two branches is to incorporate the feature map result from \normal into \secure. Particularly, the feature maps of one layer computed by \normal will be transmitted from REE to TEE and element-wise added with the feature maps generated from \secure,  which serves as the input of the subsequent layer of \secure (see \textcircled{1} in Fig.~\ref{fig:overview}). 

With such an architectural design to enhance the confidentiality of computation in TEE, it is also crucial to prevent attackers from directly leveraging the \normal to achieve satisfactory accuracy, thereby minimizing the risk of model stealing. As the current \normal is initialized with knowledge equivalent to the victim model, our subsequent objective is to transfer this knowledge from the unsecured branch to the secure branch.

\subsection{Step \textcircled{2}: Knowledge Transfer} \label{transfer}

To reduce the model confidentiality exposed in REE, \ourframework aims to transfer partial knowledge from \normal to \secure. Specifically, \normal inherits the model architecture and weights from the original victim model, encapsulating valuable ``knowledge'', in terms of discerning feature characteristics and accurate categorization. 

This knowledge transfer is optimized through the refinement of the two-branch model established in the previous step \textcircled{1}, utilizing the following objective function:
\begin{equation}\label{loss_func}
\begin{split}
L = \sum_{(x,y)} l ( f(x, \mathbf{W_R}, \mathbf{W_T}), y) + \lambda \sum_{\gamma_R,\gamma_T} g(\gamma_R + \gamma_T)
\end{split}
\end{equation}
where $x$ and $y$ denote the training input and label, while $\mathbf{W_R}$ and $\mathbf{W_T}$ represent the weights of the \normal and \secure branches, respectively. Also, f denotes the functionality of the two-branch model and l is the cross entropy loss for classification tasks.
Note that the output of \ourframework is derived from \secure, which predicts the inference result before returning it to the model users. 
During the training process, 
\ourframework transfers the knowledge and updates the weights in two branches concurrently. 
Specifically, 
The training input $x$ first runs the convolution computation with the first layer in TEE, and the resulting feature map is stored in its memory. Subsequently, the same training input $x$ is processed through the first convolution layer in REE. The intermediate results from this computation are transferred to TEE and combined element-wise with the previously stored results (see \textcircled{2} in Fig.~\ref{fig:overview}). Such 
computation is consistently applied across all subsequent layers.
Then \ourframework calculates the loss between the results from $\mathbf{W_R}$ and the target. 
Besides, $\gamma_R$ and $\gamma_T$ are the weights of the batch normalization (BN) layers, indicating the relative importance of each channel 
\cite{liu2017learning}. Here $g$ is the sparsity-induced penalty on these weights, which is the L1 normalization, and $\lambda$ balances these two terms. 
By minimizing $L$ in Eq.~\ref{loss_func}, we can achieve knowledge transfer, as analyzed in Sec. \ref{sec:analysis}.  This optimization process not only distributes the knowledge of the victim model into two branches, but also encourages sparsity in BN weights, laying the groundwork for the subsequent pruning. 

\subsection{Step \textcircled{3}-\textcircled{5}: Iterative Two-branch Pruning}\label{prune} 

{While the current two-branch substitution model effectively prevents attackers from observing critical data transmissions (e.g., feature maps from TEE) and their direct utilization of the \normal, its deployment efficiency is impeded by the redundant parameters. Specifically, the secure branch ($M_T$) deployed in TEE should be optimized for both small model sizes and low computational demands to reduce secure memory utilization and execution latency. Although there are some methods designed for pruning the DNN model (e.g., \cite{liu2017learning}), they are designed for serial DNNs with only one branch and it is nontrivial to simultaneously prune two branches.
}

To mitigate these challenges, we propose an iterative two-branch pruning approach, incorporating a channel pruning strategy. This strategy enhances the computational efficiency and reduces memory requirements of \secure while modifying the architecture in \normal. We outline this pruning strategy in Alg. \ref{alg}, which utilizes the sparsity-inducing regularization $g$ in Eq. \ref{loss_func}, to induce sparsity in the weights of the BN layers during the training phase. These sparse weights serve as the indicator, aiding in the identification of less important channels within the convolution layer. Therefore, the weights of BN layers for each channel in the \normal and \secure are extracted from the model at the first place in step \textcircled{3}. 

Considering that the initial model architecture of \normal is identical to \secure and the intermediate feature maps result from \normal are added with the result of \secure, we combine the weights of the BN layers of \normal and \secure in channel scales (step \textcircled{4} in Fig. \ref{fig:overview}) to create composite weights for the exact same architecture (line \ref{alg:composite_weights} in Alg. \ref{alg}). 
The rationale for adding the weights together is that both \normal and \secure provide information and contribute to the inference, necessitating consideration of the channel's importance in two branches. This weight addition aligns with the addition of the intermediate feature map, allowing the composite BN weights to effectively represent the importance of the merged feature map.
With the indication of composite weights, our framework prunes the less important channels accordingly. Following from line \ref{pre0} to line \ref{alg:model_inherit} in Alg.\ref{alg}, our method determines a threshold for pruning based on the composite weights and the preset pruning ratio. Each channel's composite weights are then compared to this threshold. A pruning mask, consisting of 1's and 0's, is created to indicate whether a channel should be retained or pruned.  This mask is applied simultaneously to both \secure and \normal, allowing for the pruning of the BN layer, convolution layer, and dense layer as required.

\ourframework combines pruning and retraining in a single iteration, i.e., once a model undergoes pruning, it is also fine-tuned to recover the accuracy lost due to the reduction in model size. In this way, \ourframework assesses the need for further pruning and refinement by comparing the accuracy decrease against a predefined budget for an acceptable accuracy drop. If the reduction in accuracy for \secure remains within the budget (i.e., $\theta_{drop}$ in Alg. \ref{alg}), \ourframework proceeds to the next round of pruning and retraining. However, if the accuracy drop exceeds $\theta_{drop}$, the process halts and reverts to the prior state that satisfies the accuracy criteria.

\begin{algorithm}
\caption{Iterative Two-branch Pruning}
\label{alg}
\begin{algorithmic}[1]
\REQUIRE \secure, \normal, accuracy drop budget $\theta_{drop}$, pruning ratio $p$

\WHILE{Accuracy drop $< \theta_{drop}$} \label{accloss}
    \STATE $N \gets\  $ Total number of channels in BN layers
    \STATE $BN_{T}, BN_{R} \gets\ $ BN Weights for each channel in \secure and \normal
    \STATE $BN = BN_{R} + BN_{T}$ \label{alg:composite_weights} 
    \textcolor{gray}{// Obtain composite weights}  
    \STATE $T = sort(BN)[ (N*p)]$ \label{pre0}
    \STATE Initialize pruning mask $mask[N]$ with zeros
    \FOR{$i = 1$ to $N$}
        \IF{$BN[i] > T$}
            \STATE $mask[i] \gets\ 1$
        \ENDIF
    \ENDFOR
    \STATE Prune channels in $M_T$ and $M_R$ according to $mask$\label{alg:model_inherit}
\ENDWHILE
\RETURN $M_T$, $M_R$
\end{algorithmic}
\end{algorithm}

\subsection{Step \textcircled{6}: Rollback Finalization}

Following step \textcircled{3}-\textcircled{5}, the architectural configuration of \secure remains identical to that of \normal. However, from the security perspective, it is crucial to differentiate the architecture of the \normal from the \secure, since \normal is fully exposed to the attackers (see threat model in Sec. \ref{sec:threat_model}). 
Such an architectural difference is motivated by the need to safeguard the valuable architectural IP embodied in the \ourframework, as it could potentially assist attackers in retraining to generate a high-performance model (see limitations of prior works in Sec. \ref{sec:limits_of_prior_works}).

To prevent the attacker from inferring the model architecture in TEE from that in REE, \ourframework incorporates a rollback finalization step to establish a divergence, i.e., making \secure $\ne$ \normal.
Concretely, the architecture and weights in \normal rollback specifically to the state preceding the most recent pruning iteration. Given the ample resources in REE and the iterative nature of pruning (with a small pruning ratio), the incremental increase in the model size of \normal due to the rollback introduces minimal latency overhead.
Meanwhile, the rollback of \normal to a previous state introduces additional parameters that contribute positively to overall performance. 
After receiving the feature map, 
\secure identifies and extracts the specific channel that aligns with their pre-stored feature map and runs the element-wise addition to combine the information together as the input for the next layer computation in TEE. 
This process ensures that the architecture in \secure remains confidential, effectively countering potential adversarial attempts to infer it.

\section{Experimental Evaluation}

\textbf{DNN Models.}
To validate the efficacy of \ourframework, we conduct experiments with four DNN models, including VGG18 \cite{simonyan2014very} and ResNet-20 \cite{he2016identity} models that are trained on CIFAR10 and CIFAR100 \cite{krizhevsky2010convolutional} respectively. 
In particular, for ResNet20 with skip connections, \normal is initialized from the main branch (excluding skip connections), while \secure is initialized with the original architecture.

\textbf{Hyperparameters.} During the training of the model, we utilize the SGD optimizer with a learning rate of 0.1, a momentum of 0.9, and a weight decay of 1e-4. Also, we apply a learning rate scheduler, which reduces the learning rate to one-tenth every 100 epochs. The sparsity regularization $\lambda$ in Eq. \ref{loss_func} is set to  $10^{-4}$ in our experiment. For the iterative pruning process, we define the pruning ratio ($p$) as 10\%, indicating a reduction of 10\% of the total number of channels across the entire model.

\textbf{Hardware Setup.}
We execute the implementation of \ourframework on a Raspberry Pi 3 ModelB equipped with a Broadcom BCM2837 64-bit ARM Cortex-A53 Quad-Core Processor and 1GB of RAM. For generality, we employ Open Portable TEE (OP-TEE) \cite{optee}, a versatile open-source TEE framework as the platform to verify \ourframework.

\begin{table}[tb!]
    \centering
    \resizebox{0.9\linewidth}{!}{
    \begin{tabular}{cc|cccc}
    \hline 
        \textbf{Datasets} &\textbf{DNN} & \begin{tabular}[c]{@{}c@{}}\textbf{Victim}\\\textbf{ Acc. (\%)}\end{tabular} & \begin{tabular}[c]{@{}c@{}}\textbf{\ourframework}\\ \textbf{Acc. (\%)}\end{tabular} & \begin{tabular}[c]{@{}c@{}}\textbf{Attack}\\ \textbf{Acc. (\%)}\end{tabular} & \begin{tabular}[c]{@{}c@{}} \textbf{Acc.} \\\textbf{Gap (\%)}\end{tabular}\\ \hline
         \multirow{2}{*}{CIFAR10}  & VGG18   & 91.29 & 90.72 & 69.80 & \textbf{20.92}\\
         & ResNet20 & 92.27 & 91.68  & 10.00 & \textbf{81.68} \\ \hline
         \multirow{2}{*}{CIFAR100} & VGG18 & 67.41 & 68.37 & 42.64 & \textbf{25.73}\\
         & ResNet20 & 71.03 & 69.49 & 20.29 &\textbf{48.54} \\ \hline 
    \end{tabular}}
    \caption{The performance of \ourframework and its protection against direct model usage (i.e., Attack Acc.).}
\label{training}
\end{table}

\subsection{Performance and Security Evaluation}

We comprehensively evaluate \ourframework from two perspectives: (1) \textbf{Performance}: whether \ourframework can preserve the accuracy of the original model and (2) \textbf{Security}: whether an attacker can achieve comparable accuracy using \normal that is accessible in REE. 

The experimental results in Tab.\ref{training} show that the inference accuracy \ourframework is comparable to that of the victim model. Note this is a common ``security-performance'' trade-off, since we assume the original victim model from the vendor is highly optimized (see threat model in Sec. \ref{sec:threat_model}); thus, any further pruning will introduce performance loss. Leveraging the step \textcircled{6} of \ourframework, i.e., rollback finalization, the accuracy loss can be recovered once again. It is crucial to emphasize that the accuracy loss serves as the adjustable budget for achieving secure deployment, a parameter that defenders can tailor according to their specific requirements and trade-offs.

To evaluate the security protection of \ourframework against model stealing, we compare two accuracy metrics: (1) the accuracy obtained by a benign user from the \secure output and (2) the accuracy achieved by the attacker when extracting \normal in REE and transplanting it for direct use. From Tab. \ref{training}, the achievable accuracy by the attacker is low, i.e., there is at least a 20\% accuracy gap. Such a gap is more pronounced when the victim model is a ResNet model since the unavailability of data on residual blocks significantly impedes the attacker's ability to achieve high performance in direct use.

\subsection{Protection Against Model Fine-tuning}
\begin{figure}[tb!]
  \centering
  \includegraphics[width=0.3\textwidth]{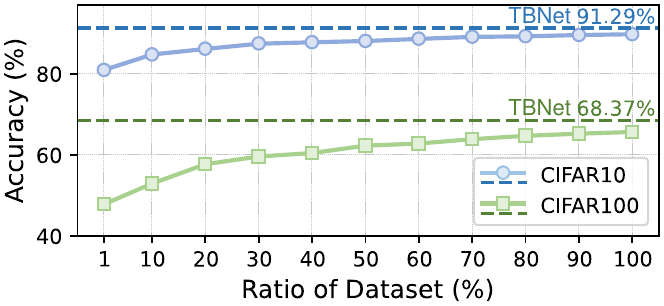}
  \caption{Accuracy of attackers fine-tuning the \normal of VGG18 under varying dataset availability.} 
  \label{fig:retraining}
\end{figure}

To demonstrate the comprehensive and robust protection of \ourframework, we assess its resilience against stronger attackers who attempt to fine-tune the model extracted from REE (\normal). Our evaluation simulates the attacker's fine-tuning process with varying training data availability, from 1\% to 100\%.
As shown in Fig.~\ref{fig:retraining}, even if the attacker fine-tunes the \normal of VGG18 with 100\% training dataset, s/he cannot achieve the performance as high as \ourframework (e.g., 65.59\% vs 68.37\% on CIFAR100). This is because: (1) the post-prune architecture  \normal is downgraded compared to the victim model and (2) the existence of \secure further enhances the overall performance of \ourframework.

\subsection{Hardware Efficiency}

Here we demonstrate the hardware efficiency of \ourframework compared to the case where victim models are fully executed in TEE (referred to as the \textbf{baseline}).

\textbf{Memory Usage}.
Secure memory usage is a main limit while deploying DNN models within TEE, which makes optimizing secure memory usage essential. In our study, we assess the memory efficiency of \ourframework by comparing the memory usage of \secure to the baseline approach. Our experiments are based on the premise that the model vendor already has a compact model, which cannot be further pruned without loss of accuracy. This condition underscores the efficacy of \ourframework, which achieves up to a 2.45x reduction in memory usage. This improvement, demonstrated using VGG18 and CIFAR10, can be attributed to the elimination of redundancies inherent in the two-branch model architecture, even if the victim model is well-pruned.
These experimental results demonstrate \ourframework's potential in generating secure and lightweight DNN on the edge.

\textbf{Inference Latency}. 
We also evaluate the performance of \ourframework by measuring the inference latency of its hardware implementations. Specifically, the baseline implements the entire victim model in TEE, while \ourframework only implements the secure branch (\secure) in TEE. We show the experimental results in Tab.\ref{exe}, which indicates that the implementation of \ourframework helps to reduce the inference latency up to 1.22x compared with the baseline method. This demonstrates that our protection method is more advantageous compared with the baseline method, making it a better solution for time-sensitive applications. 
Note that the result of latency reduction is the minimized results achievable by \ourframework, which does not apply any other optimization strategies, we discuss the optimization ways to accelerate the model inference in Sec. \ref{sec:discussion_extension}. 

\section{Discussion}\label{sec:ablation}

\begin{table}[tb!]
\centering
\begin{tabular}{c|ccc}
\hline
         & \textbf{\ourframework} & \textbf{\secure} & \textbf{Acc. Drop} \\ \hline
VGG18    &  91.29\%  &  87.57\%   &       \textbf{3.72\%}                 \\
Resnet20 &     92.27\%        &  89.41\%       &    \textbf{2.86\%}       \\ \hline
\end{tabular}
\caption{Accuracy comparison between the best possible \secure (i.e., re-training with all training data) and \ourframework.}
\label{single}
\end{table}

\subsection{Necessity of Unsecured Branch}
While \ourframework demonstrates superior performance over the existing solutions, by using the two-branch DNN model architecture,
the role of \normal in REE remains unexplored. In other words, it is unclear whether \secure alone could match the performance of \ourframework without the support and information from \normal.
To investigate the significance of \normal, we remove its architecture and weights from \ourframework and leave the \secure as the only branch of the \ourframework. Then retrain \secure with the entire training dataset to evaluate its optimal performance.
Tab. \ref{single} presents a comparison of accuracy between \ourframework and its single branch \secure, across two different model architectures, VGG18 and Resnet20, on the CIFAR10 dataset. 
Results indicate a decline in performance accuracy (up to 3.72\%) when operating without the unsecured branch (\normal), signifying that the intermediate results transmitted from REE are necessary for the overall performance.

\subsection{Analysis of Knowledge Transfer} \label{sec:analysis}

\begin{figure}[tb!]
  \centering
  \includegraphics[width=0.36\textwidth]{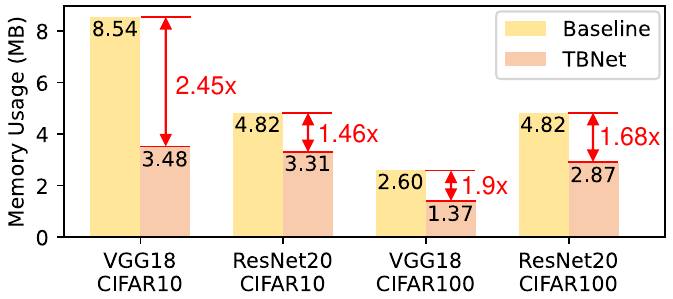}
  \caption{The comparison of memory usage in TEE.} 
  \label{fig:memory}
\end{figure}

\begin{table}[tb!]
\centering
\begin{tabular}{c|ccc}
\hline
         &\begin{tabular}[c]{@{}c@{}} \textbf{Baseline} \end{tabular} &
         \begin{tabular}[c]{@{}c@{}}\textbf{\ourframework} \end{tabular} &\begin{tabular}[c]{@{}c@{}}\textbf{Reduction}\end{tabular}  \\ \hline
VGG18   &   2.3983         & 1.9589  &  
              
\textbf{1.22$\times$}
\\
ResNet20 &     3.7425     &  3.2667 &     

\textbf{1.15$\times$}
\\ \hline
\end{tabular}
\caption{The inference latency (s) of models for CIFAR10.}

\label{exe}
\end{table}

\begin{figure}[tb!]
  \centering
  \includegraphics[width=0.37\textwidth]{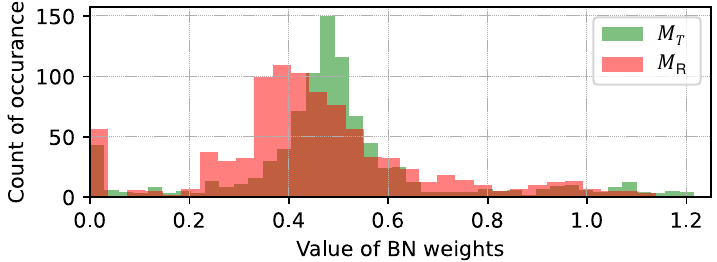}
   \caption{Distribution of BN weights in \secure and \normal after knowledge transfer.}  
  \label{fig:distribution}
\end{figure}

As discussed in Sec. \ref{transfer}, one important contribution of \ourframework is to reduce the model confidentiality exposure in REE, i.e., by transferring partial knowledge from \normal to \secure. In Fig. \ref{fig:distribution}, we visualize the distributions of BN weights in \normal and \secure, respectively, after applying the knowledge transfer. These experimental results demonstrate that the knowledge from the victim model is effectively distributed between \normal and \secure. Recalling that BN weights are used to indicate channel importance, it is noteworthy that, on average, channels in \normal demonstrate lower values compared to those in \secure, considering that channels with smaller BN weights contribute less to the model performance \cite{liu2017learning}.

\subsection{Potential Extension of \ourframework}\label {sec:discussion_extension}
The experimental results in Tab. \ref{exe} demonstrate the minimal achievable inference latency reduction by \ourframework. In real-world scenarios, the performance of \ourframework can be further enhanced by using diverse techniques, such as multiple thread execution, GPU acceleration, and specific libraries for DNN, all of which can greatly reduce the execution time of \normal in REE. In other words, our proposed \ourframework framework can be compatible with and enhanced by any existing hardware acceleration techniques. The rapid development of AI hardware accelerators highlights the great potential of our proposed two-branch methodology, which enables secure collaborative inference between REE and TEE for both performance and security. 

\section{Conclusion}
This paper presents \ourframework, a novel neural architectural defense framework to facilitate efficient DNN model protection using TEE. \ourframework generates a two-branch substitution model and deploys the secure branch within TEE for security enhancement. It also minimizes the performance overhead by offloading the unsecured branch to REE. Experimental results from diverse model architectures and datasets demonstrate that, \ourframework reduces the execution time of the model inference by 1.22$\times$, while protecting the functionality and architecture of the model against stealing even if the attacker has entire training datasets.

\section*{Acknowledgement}

This work is supported in part by the U.S. National Science Foundation under Grants OAC-2319962, CNS-2239672, CNS-2153690, CNS-2326597, and CNS-2247892. 

\input{reference.bbl}

\end{document}

%% file: reference.bbl

%% file: main.bbl
\begin{thebibliography}{21}


\ifx \showCODEN    \undefined \def \showCODEN     #1{\unskip}     \fi
\ifx \showDOI      \undefined \def \showDOI       #1{#1}\fi
\ifx \showISBNx    \undefined \def \showISBNx     #1{\unskip}     \fi
\ifx \showISBNxiii \undefined \def \showISBNxiii  #1{\unskip}     \fi
\ifx \showISSN     \undefined \def \showISSN      #1{\unskip}     \fi
\ifx \showLCCN     \undefined \def \showLCCN      #1{\unskip}     \fi
\ifx \shownote     \undefined \def \shownote      #1{#1}          \fi
\ifx \showarticletitle \undefined \def \showarticletitle #1{#1}   \fi
\ifx \showURL      \undefined \def \showURL       {\relax}        \fi
\providecommand\bibfield[2]{#2}
\providecommand\bibinfo[2]{#2}
\providecommand\natexlab[1]{#1}
\providecommand\showeprint[2][]{arXiv:#2}

\bibitem[Batina et~al\mbox{.}(2019)]%
        {batina2019csi}
\bibfield{author}{\bibinfo{person}{Lejla Batina}, \bibinfo{person}{Shivam Bhasin}, \bibinfo{person}{Dirmanto Jap}, {and} \bibinfo{person}{Stjepan Picek}.} \bibinfo{year}{2019}\natexlab{}.
\newblock \showarticletitle{$\{$CSI$\}$$\{$NN$\}$: Reverse engineering of neural network architectures through electromagnetic side channel}. In \bibinfo{booktitle}{\emph{USENIX Security'19}}.
\newblock


\bibitem[Chen et~al\mbox{.}(2019)]%
        {chen2019deepattest}
\bibfield{author}{\bibinfo{person}{Huili Chen}, \bibinfo{person}{Cheng Fu}, \bibinfo{person}{Bita~Darvish Rouhani}, \bibinfo{person}{Jishen Zhao}, {and} \bibinfo{person}{Farinaz Koushanfar}.} \bibinfo{year}{2019}\natexlab{}.
\newblock \showarticletitle{DeepAttest: An end-to-end attestation framework for deep neural networks}. In \bibinfo{booktitle}{\emph{ISCA'19}}.
\newblock


\bibitem[He et~al\mbox{.}({[n.\,d.]})]%
        {he2016identity}
\bibfield{author}{\bibinfo{person}{Kaiming He}, \bibinfo{person}{Xiangyu Zhang}, \bibinfo{person}{Shaoqing Ren}, {and} \bibinfo{person}{Jian Sun}.} \bibinfo{year}{[n.\,d.]}\natexlab{}.
\newblock \showarticletitle{Identity mappings in deep residual networks}. In \bibinfo{booktitle}{\emph{Computer Vision--ECCV 2016}}.
\newblock


\bibitem[Isakov et~al\mbox{.}(2019)]%
        {isakov2019survey}
\bibfield{author}{\bibinfo{person}{Mihailo Isakov}, \bibinfo{person}{Vijay Gadepally}, \bibinfo{person}{Karen~M Gettings}, {and} \bibinfo{person}{Michel~A Kinsy}.} \bibinfo{year}{2019}\natexlab{}.
\newblock \showarticletitle{Survey of attacks and defenses on edge-deployed neural networks}. In \bibinfo{booktitle}{\emph{HPEC'19}}.
\newblock


\bibitem[Jagielski et~al\mbox{.}(2020)]%
        {jagielski2020high}
\bibfield{author}{\bibinfo{person}{Matthew Jagielski}, \bibinfo{person}{Nicholas Carlini}, \bibinfo{person}{David Berthelot}, \bibinfo{person}{Alex Kurakin}, {and} \bibinfo{person}{Nicolas Papernot}.} \bibinfo{year}{2020}\natexlab{}.
\newblock \showarticletitle{High accuracy and high fidelity extraction of neural networks}. In \bibinfo{booktitle}{\emph{USENIX Security'20}}.
\newblock


\bibitem[Krizhevsky and Hinton(2010)]%
        {krizhevsky2010convolutional}
\bibfield{author}{\bibinfo{person}{Alex Krizhevsky} {and} \bibinfo{person}{Geoff Hinton}.} \bibinfo{year}{2010}\natexlab{}.
\newblock \showarticletitle{Convolutional deep belief networks on cifar-10}.
\newblock \bibinfo{journal}{\emph{Unpublished manuscript}} \bibinfo{volume}{40}, \bibinfo{number}{7} (\bibinfo{year}{2010}), \bibinfo{pages}{1--9}.
\newblock


\bibitem[Limited(2023)]%
        {armtrustzone}
\bibfield{author}{\bibinfo{person}{Arm Limited}.} \bibinfo{year}{2023}\natexlab{}.
\newblock \bibinfo{title}{TrustZone for Cortex-M}.
\newblock
\newblock
\urldef\tempurl%
\url{https://www.arm.com/technologies/trustzone-for-cortex-m}
\showURL{%
\tempurl}


\bibitem[Limited(2019)]%
        {optee}
\bibfield{author}{\bibinfo{person}{Linaro Limited}.} \bibinfo{year}{2019}\natexlab{}.
\newblock \bibinfo{title}{Open Portable Trusted Execution Environment}.
\newblock
\newblock


\bibitem[Liu et~al\mbox{.}(2017)]%
        {liu2017learning}
\bibfield{author}{\bibinfo{person}{Zhuang Liu}, \bibinfo{person}{Jianguo Li}, \bibinfo{person}{Zhiqiang Shen}, \bibinfo{person}{Gao Huang}, \bibinfo{person}{Shoumeng Yan}, {and} \bibinfo{person}{Changshui Zhang}.} \bibinfo{year}{2017}\natexlab{}.
\newblock \showarticletitle{Learning efficient convolutional networks through network slimming}. In \bibinfo{booktitle}{\emph{ICCV'17}}.
\newblock


\bibitem[Liu et~al\mbox{.}(2023)]%
        {liu2023mirrornet}
\bibfield{author}{\bibinfo{person}{Ziyu Liu}, \bibinfo{person}{Yukui Luo}, \bibinfo{person}{Shijin Duan}, \bibinfo{person}{Tong Zhou}, {and} \bibinfo{person}{Xiaolin Xu}.} \bibinfo{year}{2023}\natexlab{}.
\newblock \showarticletitle{MirrorNet: A TEE-Friendly Framework for Secure On-Device DNN Inference}. In \bibinfo{booktitle}{\emph{IEEE/ACM International Conference on Computer Aided Design (ICCAD)}}. IEEE, \bibinfo{pages}{1--9}.
\newblock


\bibitem[Lou et~al\mbox{.}(2022)]%
        {lou2022ownership}
\bibfield{author}{\bibinfo{person}{Xiaoxuan Lou}, \bibinfo{person}{Shangwei Guo}, \bibinfo{person}{Jiwei Li}, {and} \bibinfo{person}{Tianwei Zhang}.} \bibinfo{year}{2022}\natexlab{}.
\newblock \showarticletitle{Ownership verification of dnn architectures via hardware cache side channels}.
\newblock \bibinfo{journal}{\emph{TCSVT}} (\bibinfo{year}{2022}).
\newblock


\bibitem[Mo et~al\mbox{.}(2020)]%
        {mo2020darknetz}
\bibfield{author}{\bibinfo{person}{Fan Mo}, \bibinfo{person}{Ali~Shahin Shamsabadi}, \bibinfo{person}{Kleomenis Katevas}, \bibinfo{person}{Soteris Demetriou}, \bibinfo{person}{Ilias Leontiadis}, \bibinfo{person}{Andrea Cavallaro}, {and} \bibinfo{person}{Hamed Haddadi}.} \bibinfo{year}{2020}\natexlab{}.
\newblock \showarticletitle{Darknetz: towards model privacy at the edge using trusted execution environments}. In \bibinfo{booktitle}{\emph{MobiSys'20}}.
\newblock


\bibitem[Mo et~al\mbox{.}(2022)]%
        {mo2022sok}
\bibfield{author}{\bibinfo{person}{Fan Mo}, \bibinfo{person}{Zahra Tarkhani}, {and} \bibinfo{person}{Hamed Haddadi}.} \bibinfo{year}{2022}\natexlab{}.
\newblock \showarticletitle{SoK: machine learning with confidential computing}.
\newblock \bibinfo{journal}{\emph{arXiv preprint arXiv:2208.10134}} (\bibinfo{year}{2022}).
\newblock


\bibitem[Murshed et~al\mbox{.}(2021)]%
        {murshed2021machine}
\bibfield{author}{\bibinfo{person}{MG~Sarwar Murshed}, \bibinfo{person}{Christopher Murphy}, \bibinfo{person}{Daqing Hou}, \bibinfo{person}{Nazar Khan}, \bibinfo{person}{Ganesh Ananthanarayanan}, {and} \bibinfo{person}{Faraz Hussain}.} \bibinfo{year}{2021}\natexlab{}.
\newblock \showarticletitle{Machine learning at the network edge: A survey}.
\newblock \bibinfo{journal}{\emph{ACM Computing Surveys (CSUR)}} \bibinfo{volume}{54}, \bibinfo{number}{8} (\bibinfo{year}{2021}), \bibinfo{pages}{1--37}.
\newblock


\bibitem[Simonyan and Zisserman(2014)]%
        {simonyan2014very}
\bibfield{author}{\bibinfo{person}{Karen Simonyan} {and} \bibinfo{person}{Andrew Zisserman}.} \bibinfo{year}{2014}\natexlab{}.
\newblock \showarticletitle{Very deep convolutional networks for large-scale image recognition}.
\newblock \bibinfo{journal}{\emph{arXiv preprint arXiv:1409.1556}} (\bibinfo{year}{2014}).
\newblock


\bibitem[Sun et~al\mbox{.}({[n.\,d.]})]%
        {sun2023shadownet}
\bibfield{author}{\bibinfo{person}{Zhichuang Sun}, \bibinfo{person}{Ruimin Sun}, \bibinfo{person}{Changming Liu}, \bibinfo{person}{Amrita~Roy Chowdhury}, \bibinfo{person}{Long Lu}, {and} \bibinfo{person}{Somesh Jha}.} \bibinfo{year}{[n.\,d.]}\natexlab{}.
\newblock \showarticletitle{Shadownet: A secure and efficient on-device model inference system for convolutional neural networks}. In \bibinfo{booktitle}{\emph{SP'23}}.
\newblock


\bibitem[Sun et~al\mbox{.}(2021)]%
        {sun2021mind}
\bibfield{author}{\bibinfo{person}{Zhichuang Sun}, \bibinfo{person}{Ruimin Sun}, \bibinfo{person}{Long Lu}, {and} \bibinfo{person}{Alan Mislove}.} \bibinfo{year}{2021}\natexlab{}.
\newblock \showarticletitle{Mind your weight (s): A large-scale study on insufficient machine learning model protection in mobile apps}. In \bibinfo{booktitle}{\emph{USENIX Security'21}}.
\newblock


\bibitem[VanNostrand et~al\mbox{.}(2019)]%
        {vannostrand2019confidential}
\bibfield{author}{\bibinfo{person}{Peter~M VanNostrand}, \bibinfo{person}{Ioannis Kyriazis}, \bibinfo{person}{Michelle Cheng}, \bibinfo{person}{Tian Guo}, {and} \bibinfo{person}{Robert~J Walls}.} \bibinfo{year}{2019}\natexlab{}.
\newblock \showarticletitle{Confidential deep learning: Executing proprietary models on untrusted devices}.
\newblock \bibinfo{journal}{\emph{arXiv preprint arXiv:1908.10730}} (\bibinfo{year}{2019}).
\newblock


\bibitem[Wang et~al\mbox{.}(2023)]%
        {IP_Protection_TinyML}
\bibfield{author}{\bibinfo{person}{Jinwen Wang}, \bibinfo{person}{Yuhao Wu}, \bibinfo{person}{Han Liu}, \bibinfo{person}{Bo Yuan}, \bibinfo{person}{Roger Chamberlain}, {and} \bibinfo{person}{Ning Zhang}.} \bibinfo{year}{2023}\natexlab{}.
\newblock \showarticletitle{IP Protection in TinyML}. In \bibinfo{booktitle}{\emph{DAC}}.
\newblock


\bibitem[Xiao et~al\mbox{.}(2019)]%
        {xiao2019fast}
\bibfield{author}{\bibinfo{person}{Xueli Xiao}, \bibinfo{person}{Thosini~Bamunu Mudiyanselage}, \bibinfo{person}{Chunyan Ji}, \bibinfo{person}{Jie Hu}, {and} \bibinfo{person}{Yi Pan}.} \bibinfo{year}{2019}\natexlab{}.
\newblock \showarticletitle{Fast deep learning training through intelligently freezing layers}. In \bibinfo{booktitle}{\emph{iThings'19}}.
\newblock


\bibitem[Zhang et~al\mbox{.}(2022)]%
        {zhang2022teeslice}
\bibfield{author}{\bibinfo{person}{Ziqi Zhang}, \bibinfo{person}{Lucien~KL Ng}, \bibinfo{person}{Bingyan Liu}, \bibinfo{person}{Yifeng Cai}, \bibinfo{person}{Ding Li}, \bibinfo{person}{Yao Guo}, {and} \bibinfo{person}{Xiangqun Chen}.} \bibinfo{year}{2022}\natexlab{}.
\newblock \showarticletitle{Teeslice: slicing dnn models for secure and efficient deployment}. In \bibinfo{booktitle}{\emph{AISTA'22}}.
\newblock


\end{thebibliography}
